\documentclass[preprint,12pt]{elsarticle}

\usepackage{amssymb}
\usepackage{amsmath}
\usepackage[utf8]{inputenc}

\usepackage[para,online,flushleft]{threeparttable}
\usepackage[english]{babel}
\usepackage{amsthm}
\usepackage{amsmath}
\usepackage{varwidth}
\usepackage{subcaption}
\usepackage{calc}
\usepackage{blindtext}
\usepackage{multicol}
\usepackage{graphicx}
\usepackage{paralist}
\usepackage[english]{babel}
\usepackage{float}
\usepackage{makecell}
\usepackage{multirow}
\usepackage{amsfonts}
\usepackage{booktabs}
\usepackage{siunitx}
\usepackage{etoolbox}
\usepackage{numprint}
\usepackage[T1]{fontenc}
\usepackage{algorithm}
\usepackage{flushend}
\usepackage[noend]{algpseudocode}
\usepackage{textcomp}
\usepackage{xcolor}
\usepackage{fancyvrb}
\usepackage{paralist}
\usepackage[inline]{enumitem}
\usepackage{booktabs}
\usepackage{color}
\usepackage{colortbl}
\usepackage{flexisym}
\usepackage{array}
\usepackage{xparse}
\usepackage{xspace}
\usepackage[flushleft]{threeparttable}
\usepackage[nobreak]{mdframed}
\usepackage{framed}
\usepackage{pifont}% http://ctan.org/pkg/pifont
\usepackage{balance}

\newcommand{\cmark}{\ding{51}}%
\newcommand{\xmark}{\ding{55}}%

\newcommand{\ie}{\hbox{\emph{i.e.}}\xspace}

\newcommand{\etc}{\hbox{\emph{etc.}}\xspace}

\newcommand{\etal}{\hbox{\emph{et al.}}\xspace}

\usepackage{listings}
\usepackage[]{xcolor}
\theoremstyle{definition}
\newtheorem{definition}{Definition}
\newcommand{\toolname}{\textsc{Mokav}\xspace}
\newcommand{\initprompt}{\textsc{pr0}\xspace}
\newcommand{\iterprompt}{\textsc{pr1}\xspace}
\newcommand{\pynguin}{\textsc{Pynguin}\xspace}
\newcommand{\dpp}{\textsc{DPP}\xspace}
\newcommand{\mainds}{\textsc{C4DET}\xspace}

\newcommand{\maindssize}{1,535\xspace}
\newcommand{\qbsize}{32\xspace}
\newcommand{\maindsproblems}{221\xspace}
\newcommand{\mycomment}[1]{}

\usepackage[]{xcolor}
\newcommand{\TODO}[1]{\textcolor{red}{#1}\GenericWarning{}{LaTeX Warning: TODO: #1}}\newcommand\todo\TODO

\newcommand{\MODIFIED}[1]{\textcolor{black}{#1}}\newcommand\modified\MODIFIED
\newcommand{\MODIFIEDTWO}[1]{\textcolor{black}{#1}}\newcommand\modifiedtwo\MODIFIEDTWO
\newcommand{\MODIFIEDTHREE}[1]{\textcolor{black}{#1}}\newcommand\modifiedthree\MODIFIEDTHREE

\newcommand{\GHilight}{\makebox[0pt][l]{\color{green!20}\rule[-4pt]{1\linewidth}{10pt}}}

\newcommand{\RHilight}{\makebox[0pt][l]{\color{pink}\rule[-4pt]{1\linewidth}{10pt}}}

\algnewcommand{\Inputs}[1]{%
  \Statex \textbf{Inputs:}
  \Statex \hspace*{\algorithmicindent}\parbox[t]{.8\linewidth}{\raggedright #1}
}
\algnewcommand{\Outputs}[1]{%
  \Statex \textbf{Outputs:}
  \Statex \hspace*{\algorithmicindent}\parbox[t]{.8\linewidth}{\raggedright #1}
}
\algnewcommand{\Initialize}[1]{%
  \State \textbf{Initialize:}
  \Statex \hspace*{\algorithmicindent}\parbox[t]{.8\linewidth}{\raggedright #1}
}

\definecolor{javared}{rgb}{0.6,0,0} % for strings
\definecolor{javagreen}{rgb}{0.25,0.5,0.35} % comments
\definecolor{javapurple}{rgb}{0.5,0,0.35} % keywords
\definecolor{javadocblue}{rgb}{0.25,0.35,0.75} % javadoc

\lstdefinestyle{diff}{
    escapechar=\%
}
\usepackage{hyperref}
\addto\extrasenglish{%

}

\usepackage{cleveref}

\journal{Journal of Systems and Software}

\begin{document}

\begin{frontmatter}

%% Title, authors and addresses

%% use the tnoteref command within \title for footnotes;
%% use the tnotetext command for theassociated footnote;
%% use the fnref command within \author or \affiliation for footnotes;
%% use the fntext command for theassociated footnote;
%% use the corref command within \author for corresponding author footnotes;
%% use the cortext command for theassociated footnote;
%% use the ead command for the email address,
%% and the form \ead[url] for the home page:
%% \title{Title\tnoteref{label1}}
%% \tnotetext[label1]{}
%% \author{Name\corref{cor1}\fnref{label2}}
%% \ead{email address}
%% \ead[url]{home page}
%% \fntext[label2]{}
%% \cortext[cor1]{}
%% \affiliation{organization={},
%%             addressline={},
%%             city={},
%%             postcode={},
%%             state={},
%%             country={}}
%% \fntext[label3]{}

\title{Mokav: Execution-driven Differential Testing with LLMs}

\author[eth,kth]{Khashayar Etemadi\corref{correspondingauthor}\fnref{samecontr}}
\ead{khaes@kth.se}

\author[sharif]{Bardia Mohammadi\fnref{samecontr}}
\ead{bardia.mohammadi@sharif.edu}

\author[eth]{Zhendong Su}
\ead{zhendong.su@inf.ethz.ch}

\author[kth]{Martin Monperrus}
\ead{monperrus@kth.se}

\address[eth]{ETH Zurich, Switzerland}
\address[kth]{KTH Royal Institute of Technology, Sweden}
\address[sharif]{Sharif University of Technology, Iran}

\cortext[correspondingauthor]{Corresponding authors.}
\fntext[samecontr]{Both authors contributed equally to this research.}

%% Abstract
\begin{abstract}
It is essential to detect functional differences between programs in various software engineering tasks, such as automated program repair, mutation testing, and code refactoring. The problem of detecting functional differences between two programs can be reduced to searching for a difference exposing test (DET): a test input that results in different outputs on the subject programs. 
In this paper, we propose \toolname, a novel execution-driven tool that leverages LLMs to generate DETs. \toolname takes two versions of a program (P and Q) and an example test input. When successful, \toolname generates a valid DET, a test input that leads to provably different outputs on P and Q. \toolname iteratively prompts an LLM with a specialized prompt to generate new test inputs. At each iteration, \toolname provides execution-based feedback from previously generated tests until the LLM produces a DET.
We evaluate \toolname on \maindssize pairs of Python programs collected from the Codeforces competition platform and \qbsize pairs of programs from the QuixBugs dataset. Our experiments show that \toolname outperforms the state-of-the-art, Pynguin and Differential Prompting, by a large margin. \toolname can generate DETs for 81.7\% (1,255/\maindssize) of the program pairs in our benchmark (versus 4.9\% for Pynguin and 37.3\% for Differential Prompting). We demonstrate that the iterative and execution-driven feedback components of the system contribute to its high effectiveness.
\end{abstract}

%%Research highlights
%\begin{highlights}
%\item Mokav is the first LLM-based tool for difference exposing test generation
%\item C4DET is a curated dataset of 1,535 pair of programs with small semantical differences
%\item The iterative execution-driven approach of Mokav outperforms state-of-the-art in differential testing
%\item An example test that guides the LLM about input structure significantly improves the effectiveness
%\end{highlights}

%% Keywords

\end{frontmatter}

%\linenumbers

\section{Introduction}
\label{sec:intro}
In differential testing, two versions of a program are given (P \& Q), and a test input is searched for such that P \& Q produce different outputs~\cite{gulzar2019perception}. This test exposes a functional difference between P \& Q, thus is called a difference exposing test (DET). \modifiedtwo{DETs are useful in various software development tasks, such as for explaining changes~\cite{castellano2022explaining}, detecting anomalies and bugs~\cite{jarman2020program,gulzar2019perception}, analyzing refactorings~\cite{daniel2007automated}.} Generating a DET is significantly more challenging than random test generation because it requires exploring the vast input space of P \& Q for the rare inputs that trigger the functional differences~\cite{mckeeman1998differential}.

% related work
Existing techniques to find difference exposing tests limit the search scope using symbolic execution~\cite{rutledge2022automating}, type aware mutation~\cite{liu2024your}, and code coverage optimization with genetic algorithms~\cite{da2020detecting}. These techniques do not leverage recent advanced methods for understanding P \& Q's semantics to guide the search for DETs. Specifically, large language models (LLMs) have shown strong performance in program comprehension~\cite{yuan2023evaluating}. This makes LLMs an interesting candidate for difference exposing test search.
Previous studies have employed LLMs to conduct various software testing tasks, such as test completion~\cite{nie2023learning}, test input generation~\cite{xia2024fuzz4all}, test oracle generation~\cite{dinella2022toga}, unit test generation~\cite{schafer2023empirical}, and GUI testing~\cite{liu2024make}. 
%Li~\etal~\cite{li2023nuances} (ASE 2023) also show that LLMs, like ChatGPT, can be used to find test inputs in specific parts of the input space; namely, fault detecting test inputs. 
To the best of our knowledge, there is no prior work on leveraging LLMs for differential testing.

In this paper, we propose \toolname, a novel LLM-based tool for differential testing. \toolname iteratively requests an LLM to generate a DET based on two programs P \& Q. At each iteration, \toolname provides three pieces of information in its prompt to the LLM. First, an example test input that produces the same output on P \& Q. This example test hints at the model regarding the type and structure of inputs acceptable by P \& Q. 
Second, \toolname runs the example test on P \& Q and adds their identical output to the prompt. The output produced by P \& Q on the example test provides extra information about their functionality. Third, \toolname collects variable values while running the example test on P \& Q and adds detected disparities in the prompt. Using this fine-grained execution data, \toolname steers the LLM towards parts of the input space that contain DETs.

We evaluate \toolname on two datasets: \qbsize pairs of Python programs in the widely used QuixBugs~\cite{lin2017quixbugs} dataset, and our own crafted dataset \mainds. We create \mainds by carefully selecting \maindssize pairs of Python programs from
% ZS: no extra space before \footnote
Codeforces\footnote{\url{https://codeforces.com/}} competitions, with a guaranteed functional difference. Our evaluation results show that \toolname outperforms our two strong baselines, Pynguin~\cite{lukasczyk2022pynguin} and Differential Prompting (\dpp)~\cite{li2023nuances}. \toolname generates a DET for 81.7\% (1,255/\maindssize) of the pairs in \mainds, while \pynguin and \dpp generate a DET for 4.9\% and 37.3\% of the pairs, respectively. This promising result indicates that \toolname effectively leverages LLMs for differential testing.
\toolname is the first LLM-based differential testing, and we provide a publicly available implementation for future research.

To summarize, we make the following contributions:
\begin{itemize}
    \item We propose a novel differential testing approach based on large-language models.     
    It consists of iteratively giving execution feedback to an LLM to direct it towards generating a test input that exposes functional differences between two programs. The approach is implemented in \toolname, made publicly available for future research: \url{https://github.com/ASSERT-KTH/mokav}    
    \item We introduce a large, carefully crafted dataset \mainds, collected from Codeforces competitions, to study differential testing in Python. \mainds contains \maindssize program pairs with guaranteed functional differences. We expect this dataset to facilitate future research on differential testing, with or without LLMs.
    \item We present a large-scale, systematic evaluation of \toolname on QuixBugs and \mainds. Our results show that \toolname successfully generates a DET for 81.7\% (1,255/\maindssize) of the pairs in \mainds, clearly outperforming the recent and strong state-of-the-art systems.
    \item We conduct an ablation study of \toolname demonstrating the impact of each component on its effectiveness. We also show that different LLMs (GPT-3.5, GPT-4o, and CodeLlama) can be successfully plugged into \toolname.
\end{itemize}

\section{Problem Statement}
\label{sec:problem_statement}

In many software engineering tasks, developers face multiple versions of a program, having to reason about their functional difference or equivalence. 
For example, developers want to assess equivalence after refactoring. Another example is when developers need to ensure the functional difference of a mutant in mutation testing.
A difference exposing test (DET) is a test input on which two given programs function differently at runtime~\cite{diffgen}, it is an existential, concrete proof of functional difference. 

\begin{definition}[Difference Exposing Test (DET)]
\label{def:det}
Let $P\colon I\to O$ and $Q\colon I\to O$ be two programs operating deterministically on the same input/output space, where $I$ is the input domain of $P$ and $Q$ and  $O$ is the output domain. Then, \emph{det} is a difference exposing test iff $det\in I \wedge P(det)\neq Q(det)$.
\end{definition}

Generating a DETs is hard. The input space of the given programs can be excessively large, which makes it impossible to check the entire input space for finding a DET. Consequently, generating a DET requires an efficient exploration of the input space, based on guesses of potential functional differences between the given programs. 
We note that the execution of programs under analysis for each sample inputs can also take time~\cite{pizzoleto2019systematic}.

Hence, an efficient DET generation tool must have two qualities: (1) leverage as much static information as possible for comprehending the behavior differences of given programs, and (2) perform a targeted input exploration to narrow down the input space before paying the price of execution. Both problems have been studied in recent years~\cite{vieirab2022semeo,schoofs2022ampyfier}, but remain open.

% \textbf{Use case.} 
%Consider automated program repair (APR) as an example. APR tools usually generate multiple patches for a given buggy program~\cite{martinez2018ultra,xia2023automated}. The users of APR tools then face this important question: ``what are the differences between these patches, and which one better repairs the bug?''. A DET for which two patches generate different outputs clearly shows the functional difference between the patches. Based on this exposed functional difference, users can decide which APR patch produces the correct output and should be considered for further assessment.

Note that per Definition \autoref{def:det}, a DET detects \emph{functional differences} between P and Q, meaning they produce different outputs for a given input. In this work, we consider functional difference to be distinct from the \emph{execution difference}.
P and Q have an execution difference if, for a given input, there is a variable \emph{v} used in both P and Q where the set of values \emph{v} takes on in P is different from the set of values \emph{v} takes on in Q. 
In essence, functional difference relates to external differences between P and Q that can be identified by an external observer; in contrast, execution difference pertains to internal differences between P and Q that are detected solely through monitoring the internal states of P and Q.
%There can be inputs for which P and Q show an execution difference, while remaining functionally the same by producing similar outputs. The execution differences captured by such inputs provide valuable behavioral information about P and Q that can be utilized for generating DETs.
The goal of this paper is to design a novel, effective approach for DET generation.

\section{Approach}
\subsection{Overview}
\label{sec:overview}

\begin{figure*}
\begin{center}
\includegraphics[width=1\textwidth]{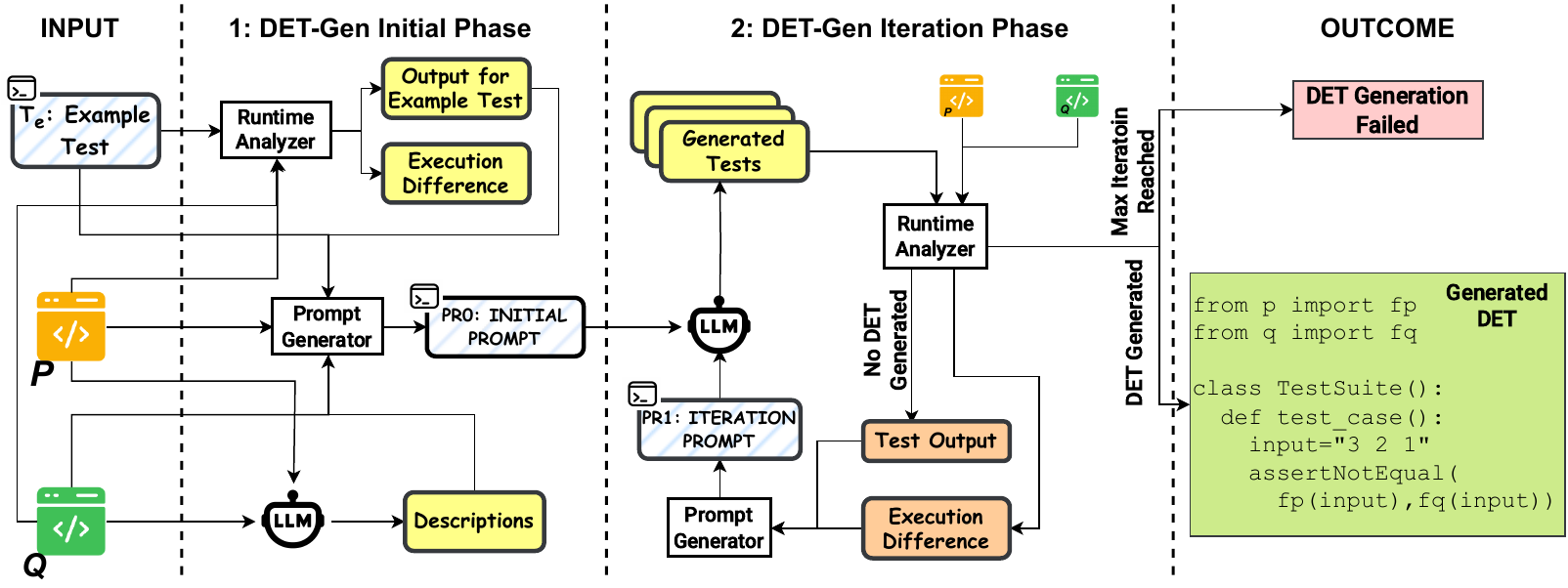}
\caption{\modifiedthree{Overview of \toolname. \toolname exposes functional differences between two programs P and Q by feeding LLMs with execution feedback.}}
\label{fig:det-gen}
\end{center}
\end{figure*}

% what we do
We design \toolname, a novel tool that uses an execution-driven approach for generating difference exposing tests (DETs) with large language models (LLMs). We implement it for generating DETs for Python programs.

% what det-gen does
\toolname takes an iterative approach to DET generation. It asks the LLM to generate a set of test inputs, gives feedback to the LLM regarding the execution data captured in the generated test inputs, and loops until a DET is generated or a maximum budget is reached.

% how det-gen works
\autoref{fig:det-gen} overviews how \toolname works. \toolname takes as input two versions of a program (P and Q) as well as a sample test input, for which both versions produce the same output. We call this input an \emph{example test}.

The example test is an essential component of \toolname for two reasons. First and foremost, \toolname utilizes the example test to extract valuable execution data. This execution data has two parts: (1) the output by P and Q, which is identical on the example test, and (2) execution differences between P and Q, if any, where an execution difference is a variable value that only occurs in one version, but does not lead to different outputs by P and Q. This detected execution difference is provided to the LLM to accurately steer it towards DET generation. Second, the example test provides valuable information for the LLM to understand the type and structure of the input space I of P and Q . \toolname's outcome is a DET, an input data for which P and Q produce different outputs.

% example of what det-gen does

%~/projs/det-gen/tmp/main/generated_tests/2346/95884
%https://chatgpt.com/share/0fea9c8e-76a0-4848-a5e1-52deaf2d5452
%https://chatgpt.com/share/9f362d78-03c9-4fbc-af55-7376ed3c6d43
\begin{lstlisting}[float,style=diff, caption={Two versions of a program that takes four numbers, sorts them, and checks if the first three or last three form a tirangle. An example test with identical output on P \& Q and a DET generated by \toolname are also shown.}, label=lst:motivating_ex]
%{\color{blue} \textbf{\textbf{Diff between P \& Q}:}}%
x = input().split()
x.sort()
for i in range(len(x)):
  x[i] = int(x[i])
x.sort()
n = 100
for j in range(2):
  if ((x[j] + x[(j + 1)]) > x[(j + 2)]):
    n = 300
  elif ((x[j] + x[(j + 1)]) == x[(j + 2)]):
    n = max(n, 200)
  else:
    n = max(n, 100)
if (n == 300):
  print('TRIANGLE')
elif (n == 200):
%\RHilight%  print('SIGMENT')
%\GHilight%  print('SEGMENT')
else:
  print('IMPOSSIBLE')
%\hrule%
%{\color{blue} \textbf{Example test:}}%
Input:
    "4 2 1 3"
Output:
    P: "TRIANGLE"
    Q: "TRIANGLE"
%\hrule%
%{\color{blue} \textbf{Generated Difference-exposing test (DET):}}%
Input:
    "5 2 1 3"
Output:
    P: "SIGMENT"
    Q: "SEGMENT"
\end{lstlisting}

\autoref{lst:motivating_ex} shows an example pair of programs given to \toolname and the DET it generates. In this example, P and Q are two real-world programs by a single user for a problem in the CodeForces\footnote{\url{https://codeforces.com/problemset/problem/6/A}} code competition. A correct program should sort the input numbers and check if the first three or the last three make up a triangle, a segment, or neither. The example test is ``4 2 1 3'', for which both versions output ``TRIANGLE''. Given P, Q, and the example test, \toolname generates a new input data: ``5 2 1 3''. On this input, P's output is ``SIGMENT'' with a misspelling, while Q correctly outputs ``SEGMENT''. This means \toolname successfully generates a DET.

% The difference is explained as follows. For the generated DET (``5 2 1 3''), both programs first sort the numbers (lines 7 and 28). Then the for loop (line 8 and 29) runs twice and at both iterations the \texttt{elif} block is executed (lines 12 and 33). Consequently, P prints ``SIGMENT'' (line 18) and Q prints ``SEGMENT'' (line 39). This generated DET arguably helps the developer to understand the behavioral difference between P and Q.

In the initial phase, \toolname creates an initial prompt based on P and Q, the example test, the execution data extracted while running the example test, and an LLM-generated description of the intention of P and Q. The description of P and Q clarifies the program intent in natural language, which is the language more familiar for many LLMs. Next, in the iteration phase, \toolname uses an LLM to generate new test inputs and iteratively gives execution based feedback to the LLM until it generates a DET. The generated DET is reported as the outcome of \toolname.

Our novel technique of generating DETs with LLMs employs two powerful features of LLMs. 
First, \toolname exploits the LLM's ability to explore the input space for a given program. As previous work demonstrates, given appropriate information, LLMs can generate a high number of diverse and novel test inputs~\cite{schafer2023empirical}. This enables us to employ LLMs for exploring different parts of the input space of P and Q.
Second, we leverage the LLM's ability to understand the semantic differences between closely related programs. Recent studies show that advanced LLMs can accurately detect inconsistencies between programs and their descriptions~\cite{li2024mutation}. This suggests that such LLMs can also accurately understand the differences between P and Q. We take advantage of this strength of LLMs to capture the functional differences between P and Q. Providing execution data to the LLM is a major part of \toolname's design to improve LLMs' understanding of P and Q.

\subsection{Input}
\label{sec:input}
The input to \toolname is a tuple in the form of $(P,Q,T_{e})$. P and Q are a pair of versions of the program for which \toolname should generate a DET. The last input to \toolname, $T_{e}$, is an example test with the same corresponding output on both P and Q, which means it does not expose functional differences. This example test is used to capture execution differences between P and Q and also to show the type of the desired test input to the LLM (see \autoref{sec:initial_phase}). In \autoref{lst:motivating_ex}, the example test hints to the model that the input contains four numbers, which makes the LLM less likely to generate invalid inputs with fewer or more numbers.

% example
% Consider \autoref{lst:motivating_ex} as an example. As explained in \autoref{sec:overview}, its goal is to check if the smallest three or largest three inputs form a triangle. When the smallest three and largest three numbers form a segment, P prints ``SIGMENT'' while, Q prints ``SEGMENT''. \toolname should generate a DET that exposes this functional difference. In \autoref{lst:motivating_ex}, the example test hints the model that the input is a string with four numbers, which makes the LLM less likely to generate invalid inputs with fewer or more numbers.

\subsection{\toolname Initial Phase}
\label{sec:initial_phase}

\begin{figure*}
\begin{center}
\includegraphics[width=0.8\textwidth]{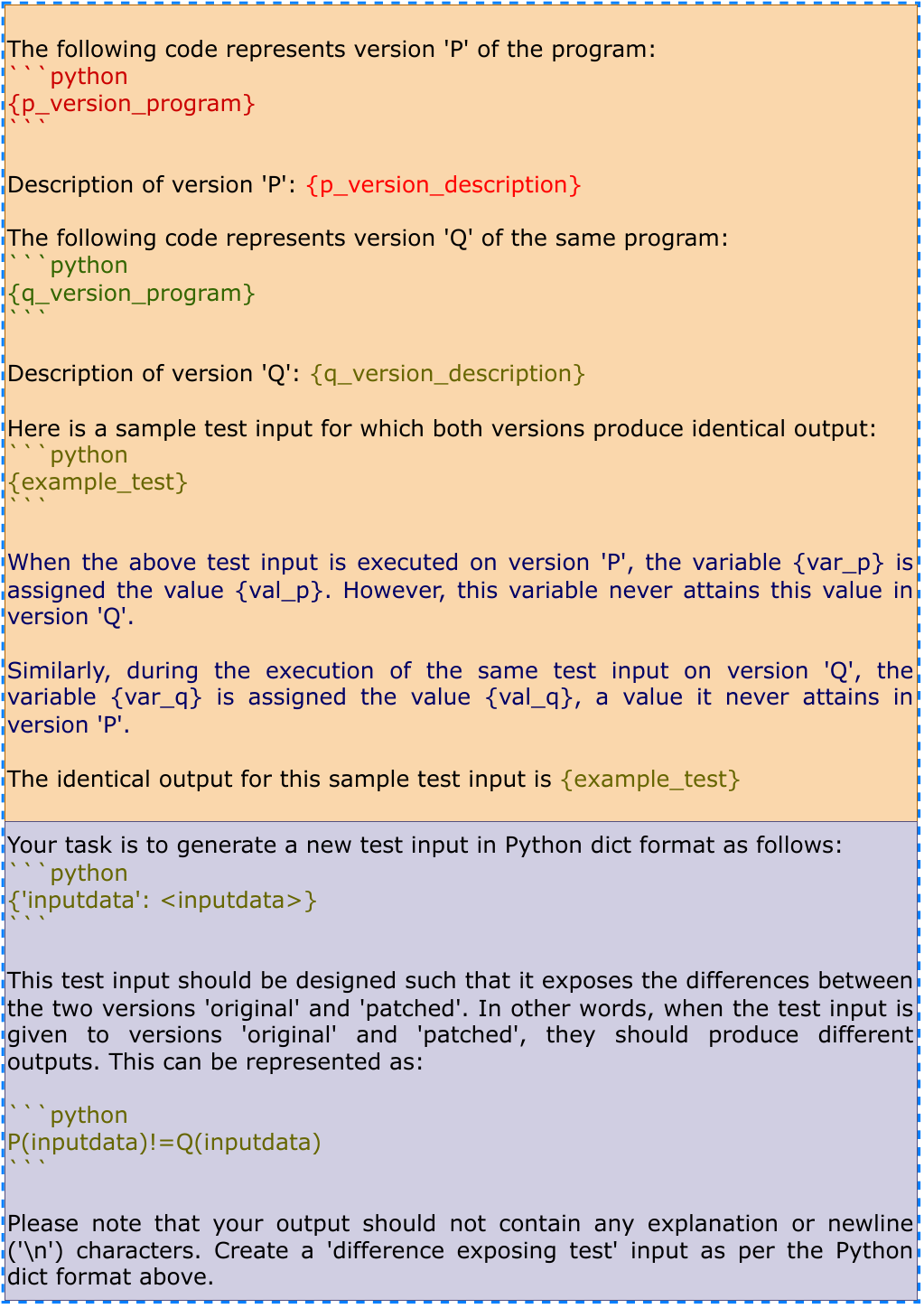}
\caption{The initial prompt (\initprompt) used in \toolname.}
\label{fig:initial_prompt}
\end{center}
\end{figure*}

In the initial phase, \toolname combines the inputs to generate an initial prompt, called \initprompt, that encompasses all the information required for starting the DET generation process.

First, \toolname generates two extra pieces of information: (1) execution data, which includes example test output and execution difference data, and (2) program descriptions. This information is obtained as follows:

\noindent
(1) \textbf{Example test output} and \textbf{execution difference}: The \emph{runtime analyzer} unit of \toolname executes the example test on P and Q to obtain the corresponding output and identify execution differences. Note that the output for the example test must be the same on both versions. If not, the generation is aborted. To identify the execution differences the runtime analyzer takes three steps as follows. First, it finds the variables that are common between P and Q. Next, it uses spotflow\footnote{\url{https://github.com/andrehora/spotflow}} to collect variable values during the execution of P and Q. Finally, the runtime analyzer identifies the first time a common variable gets a value in one of the versions, while that variable value never occurs in the other version. Such variable values are called \emph{unique variable values}~\cite{etemadi2023augmenting}, and the runtime analyzer reports them as the detected execution difference. Note that depending on the example test, the runtime analyzer may be unable to detect a unique variable value for either of the versions. In this case, that part of the data is excluded from the prompt. The output for an example test and the detected execution differences clarify the behavior of the programs, which is crucial for DET generation.

\noindent
(2) \textbf{Program descriptions:} \toolname prompts the LLM to generate natural language descriptions for P and Q. Recent studies show advanced LLMs are able to understand inconsistencies between programs and their natural language descriptions~\cite{li2024mutation}. This suggests that adding the generated descriptions to \initprompt can help the LLM to detect behavioral differences between P and Q. The prompt \toolname uses for description generation is a single line message ``\texttt{What is the intention of this code?}'', followed by the program.

The \initprompt follows the structure shown in \autoref{fig:initial_prompt}.
The main part of the prompt consists of (1) the P and Q versions of the program, (2) the description of P and Q, (3) the example test and its corresponding output, (4) the execution difference detected while running the example test, and (5) text to specify the kind of response that is expected. For the latter, the expected format of the generated test is presented. Then, the prompt defines what a difference exposing test is. To ensure the LLM understands the definition, it is written in formal terms as well as natural language. The formal definition (``\texttt{P(inputdate)!=Q(inputdata}'') helps the model relate the DET generation task with the given P and Q programs. 

From the initial prompt, the LLM generates a first candidate difference exposing test. \toolname executes both P and Q on the generated input to verify the presence of a functional difference.
If that happens, \toolname generates a DET with the \initprompt directly.

\subsection{\toolname Iteration Phase}
\label{sec:iteration_phase}

\toolname iteratively prompts the LLM until either it generates a DET or the maximum number of iterations is reached. At each iteration, after the model replies with a set of generated tests, the tests are executed on P and Q. If a test produces different outputs on P and Q, it is a valid DET, \toolname has succeeded, and we exit.
Otherwise, \toolname tries again by requesting the LLM with an iteration prompt, called \iterprompt. To create \iterprompt, \toolname selects the first generated test and uses its runtime analyzer unit to produce execution based feedback for the selected test. This feedback contains the output of P and Q on the selected test and the execution difference detected during the execution of this test (see \autoref{sec:initial_phase} for the execution difference detection description). \toolname gives feedback only on the first generated test to avoid exceeding the prompt size limit.

\modifiedtwo{At each iteration, \toolname invokes the LLM as a continuation of the initial prompt and previous iterations' prompts. This means that the request to LLM in the \textit{i}th iteration $it_{i}$ also contains the prompts from the $it_{1},it_{2},...,it_{i-1}$ iterations and the initial prompt. Therefore, the LLM leverages information regarding all previous attempts as well as the example test and descriptions at each iteration.}

%The \iterprompt prompt follows the structure shown in \autoref{fig:iteration_prompt}. The system message in \iterprompt is the same as \initprompt, thus not shown in \autoref{fig:iteration_prompt}.

The \iterprompt prompt has three parts. The first part is the detected execution difference. If the runtime analyzer does not detect an execution difference, this part is excluded for \iterprompt. The second part informs the LLM that P and Q produce the same output for the previously generated test, while we expect a test with different outputs on P and Q. The last part of the \iterprompt directly asks for another test.

At the end of the iteration phase, the result is either a generated test that exposes the functional difference or a failure message that the maximum number of iterations has been reached. 

\subsection{Outcome of \toolname}
\label{sec:ouptut}

The generated difference exposing test is presented to the user in the form of a unit test class. This test case imports P and Q as two functions \texttt{fp} and \texttt{fq}. Then a test case method passes the generated DET to \texttt{fp} and \texttt{fq} and asserts that their outputs are not equal (for example, see the outcome section of \autoref{fig:det-gen}).

\subsection{Implementation}
\label{sec:implementation}
\toolname is implemented in Python, and can be configured with different LLMs. By default, it supports OpenAI's \texttt{GPT} models and \texttt{CodeLLama-Instruct}. It connects to the \texttt{CodeLLama-Instruct} API through the HuggingFace framework, which makes it easy to connect \toolname to many other open-source models as well.

\toolname uses three important values that can be configured for each execution: the maximum number of iterations, the number of samples requested from the LLM at each invocation, and the temperature of the LLM. The default configuration is 10 iterations, 10 samples, and temperature=1.

\section{Experimental Methodology}
\label{sec:methodology}

\subsection{Research Questions}
\label{rqs}

\newcommand\rqone{How effective is \toolname for generating difference exposing tests?}

\newcommand\rqtwo{How does each component of \toolname impact its effectiveness?}

\newcommand\rqthree{How does the effectiveness of \toolname relate to the characteristics of the input programs?}

In this paper, we study the following research questions:
\begin{itemize}
    \item \textbf{RQ1} (effectiveness): \rqone \ We run \toolname on QuixBugs~\cite{lin2017quixbugs} and a large dataset of pairs of programs with functional differences collected from Code4Bench~\cite{majd2019code4bench}. We assess the effectiveness of \toolname in DET generation and compare it against two baselines: (1) a state-of-the-art search-based test generation tool, called \pynguin~\cite{lukasczyk2022pynguin}, and (2) \dpp~\cite{li2023nuances}.
    
    \item \textbf{RQ2} (ablation study): \rqtwo \ We evaluate whether and to what extent the different parts of \toolname impact its effectiveness.
    
    \item \textbf{RQ3} (impact of P and Q characteristics): \rqthree \ We check the effectiveness of \toolname in DET generation for different types of programs. More specifically, we check how the similarity between the two versions of the programs, the complexity of the test, and the length of the programs affect the effectiveness of \toolname.
\end{itemize}

\subsection{Datasets}
\label{sec:dataset}

We employ two datasets in our experiments. First, QuixBugs~\cite{lin2017quixbugs} is a dataset of 40 pairs of programs in Python that implement common programming problems, such as the greatest common divisor. For each of the 40 pairs of programs in QuixBugs, one version is buggy and the other is fixed. There is at least one input on which the buggy version either faces runtime error or has a return value different from the fixed version. This guarantees the functional difference between the two versions, with a significant functional overlap. This means that a DET is not easy to find. In this paper, we consider \qbsize program pairs from QuixBugs. For these pairs, QuixBugs contains a test that produces the same output on the buggy and fixed versions. This test should be used as the example test in \toolname's workflow.

As a second dataset used in our evaluation, we create a curated dataset of pairs of programs, which we call \mainds. \mainds contains pairs of similar programs with guaranteed functional differences. We create \mainds by carefully collecting and curating submissions to the Codeforces code competition website. We use Code4Bench~\cite{majd2019code4bench} to extract the required information for \mainds. Code4Bench contains problems, authors, submissions, and tests in Codeforces. Each submission is labeled with a verdict, which shows whether it passes the tests for the corresponding problem, gives a wrong answer, causes runtime error, causes timeout, or causes memory limit. \mainds consists of a subset of the programs and their corresponding problems and test cases in Code4Bench. These programs are selected as follows: for every author $a$ and every problem $pr$ in Codeforces, we add $\langle P,Q \rangle$ to \mainds iff they meet the following conditions:
\begin{enumerate}
    \item $P$ and $Q$ are both Python submissions by $a$ for $pr$. \modifiedtwo{$P$ and $Q$ are Python programs that do not use third-party libraries and are fully executable without manual environment setup.}
    \item $P$ is the last submission by $a$ for $pr$ that gives a wrong answer for a test and $Q$ is the first submission by $a$ for $pr$ that passes all tests. This condition ensures two important conditions: (1) there is a functional difference between $P$ and $Q$, and (2) $P$ and $Q$ are closest submissions by $a$ for $pr$ that have a guaranteed functional difference.
    \item There is at least one test input that produces the same output on $P$ and $Q$. We need this test in \mainds to use it as the example test.
    \item $P$ and $Q$ are both shorter than 2,500 tokens when tokenized by the gpt-3.5 tokenizer. This condition is needed as the context size of LLMs is limited and excessively long programs cannot be processed by them.
    \item The tests for $pr$ in Codeforces are shorter than 100 tokens when tokenized by the gpt-3.5 tokenizer. Again, this is needed as the context size of LLMs is limited, and long tests cannot be used as examples in the prompt.
    \item There are at least two submissions in \mainds for $pr$ that pass all the tests. We add this condition to have at least two reference versions to be used for oracle assessment~\cite{liu2024llm}. 
\end{enumerate}

The programs in Code4Bench are Python scripts that read inputs from \texttt{stdin} and print outputs to \texttt{stdout}. As many traditional test generation tools, such as \pynguin~\cite{lukasczyk2022pynguin}, generate tests for functions, we transform all programs to their function-based versions in \mainds. \modified{The transformation modifies input/outputs in two steps. First, it replaces \texttt{stdin} inputs with elements of an arguments array given to the function as \texttt{*args}. Secondly, it adds all the printed strings to newly defined \texttt{return\_list} array. The original print statements are removed and the function returns the \texttt{return\_list} array.} \autoref{lst:function_based} shows an example of this transformation. The source in Code4Bench is a Python script that reads a string (line 3) and prints the output (lines 6 and 8). The transformed version is a Python function that receives a list of arguments in \texttt{*args} (line 11) and creates a \texttt{return\_list} (line 12) that is returned as the output of the function (line 21). The first input is read from \texttt{args} (line 15) and the outputs are added to the \texttt{return\_list} at lines 18 and 20. This transformed version keeps the order of inputs and outputs as well as the functionality of the original source. At the same time, the transformed version is a function-based program suited for test generation without input/output.

\modified{Overall, \mainds contains \maindssize pairs of programs for \maindsproblems problems.} The longest pair of programs in \mainds has 4,242 tokens when tokenized by the gpt-3.5 tokenizer. Many advanced LLMs, such as gpt-3.5, have a 16,000 token limit. Thus, the maximum number of tokens in \mainds program pairs is low enough to be below this limit and leave space for other components of our prompt (see \autoref{fig:initial_prompt}).
The median number of tokens in \mainds pairs of programs is 248. This is a reasonable number of tokens for a Python function. It is also close to the median number of tokens in Code4Bench, meaning that \mainds is representative of programs in online competitions, which are used in many software engineering research projects~\cite{majd2019code4bench,hendrycks2021measuring}.

% \autoref{fig:submissions-cnt} shows the number of submissions per problem in \mainds. The median number of submissions per problem is eight and the minimum is two. This indicates that the problems considered in \mainds are solved by multiple users.

\begin{lstlisting}[float=tb, style=diff, caption={A program in \mainds and its function-based transformed version.}, label=lst:function_based]
%{\color{blue} \textbf{\textbf{Source of program P in Code4Bench}:}}%
reg = re.compile('(h)+(e)+(l)+(l)+(o)+')
%\RHilight%s1 = input()
li = reg.findall(s1)
if (not li):
%\RHilight%    print('NO')
else:
%\RHilight%    print('YES')
%\hrule%
%{\color{blue} \textbf{\textbf{Transformed version of P as a function-based program}:}}%
%\GHilight%def P_func(*args):
%\GHilight%    return_list = []

    reg = re.compile('(h)+(e)+(l)+(l)+(o)+')
%\GHilight%    s1 = args[0]
    li = reg.findall(s1)
    if (not li):
%\GHilight%        return_list.append('NO')
    else:
%\GHilight%        return_list.append('YES')
%\GHilight%    return return_list
\end{lstlisting}

\subsection{Baselines}
\label{sec:baselines}
We compare \toolname against two baselines: \pynguin~\cite{lukasczyk2022pynguin} and Differential Prompting (DP)~\cite{li2023nuances}.

\pynguin is a traditional test generation tool that generates random unit tests for Python programs. \pynguin employs evolutionary algorithms~\cite{panichella2017automated} to generate regression tests that maximize a fitness function, such as branch coverage. The generated tests call methods of a target module with correctly typed arguments and contain assertions about the methods' return values. The assertions check the return values with simple types, \ie, int, str, bytes, bool, and None as well as collections of simple types. We consider \pynguin in our evaluation as it is a well-documented tool and recognized as state-of-the-art in test generation for Python~\cite{li2023nuances,liu2024llm}.

The second baseline that we consider is DP~\cite{li2023nuances}. DP is an LLM-based tool that gets a buggy program and generates fault-inducing tests for it. DP works in three steps. First, it asks the LLM to generate multiple fixed versions for the buggy program. Next, it requests the LLM to produce a set of test cases for the buggy version. Finally, it runs the generated tests on the buggy and fixed versions. If the generated tests produce different outputs on buggy and fixed programs, they are reported as fault-inducing tests.

\modifiedtwo{Note that DP differs from \toolname in two major perspectives. First, the main goal of DP is to generate a test that detects the fault in a single given buggy program. In contrast, \toolname aims to generate a test that exposes functional differences between two different programs. This key difference between the goals of \toolname and DP leads to various differences between their prompting strategies. Namely, DP prompts only contain information about the buggy program, while \toolname includes information about both versions of the program. Second, \toolname adopts an iterative approach that directs the LLM toward DET generation with execution-based feedback, while DP calls the LLM only once and does not provide execution data in the prompt. Given these major differences, we cannot directly compare \toolname with DP. Instead, we isolate DP's last two steps and call it \dpp (standing for DP-Prime).} Similar to DP, \dpp works in one pass and generates all the requested tests in that one single iteration. The prompt is taken unmodified from the replication package of DP~\cite{li2023nuances}.

\subsection{Protocol RQ1 (effectiveness)}
\label{sec:rq1_protocol}
To answer \textbf{RQ1}, we run \toolname, \pynguin, and \dpp on QuixBugs and \mainds. \modified{We use the transformed version of all program pairs in \mainds for our experiments (see \autoref{sec:dataset}).}

As \pynguin is designed for generating regression tests for one program, it cannot be directly used for DET generation. To address this challenge, we use \pynguin in three steps. For a given pair of programs $\langle P, Q \rangle$, we first run \pynguin 10 times on version P of the program, each time with a time budget of 1,000 seconds. Second, we run the generated tests on P and Q and record their outputs. Finally, if one of the generated tests produces different results on P and Q, we report it as a DET generated by \pynguin. Using this method, we can measure \pynguin's effectiveness in DET generation.

We run \toolname for 10 iterations and generate 10 samples at each iteration. We also generate 100 samples in \dpp's single iteration. For this experiment, we select \texttt{gpt-3.5-turbo-0125} as the LLM in use, as it provides a powerful and affordable model. To take advantage of gpt-3.5's ability for generating diverse responses, we use temperature=1 in this experiment.

For each tool T, we run it on Quixbugs and \mainds and measure three metrics.
First, we count the number of $\langle P, Q \rangle$ pairs for which T generates a DET, and call it $\#Success\_Pairs$. This number is the main metric to compare the DET generation effectiveness of the considered tools. To better understand the effect of \toolname's iterative approach, we measure this main metric for each of \toolname iterations. If \toolname generates DETs for more pairs after its first iteration, it indicates the effectiveness of the iterative approach.
%We also compute the overlap between $<P,Q>$ pairs for which each tool generates a DET, to assess whether they can replace each other.
Second, we measure $\#TT$, which is the total number of tests generated by T. $\#TT$ indicates the scale of our study. 
Third, we measure $DET\_Pr$, the number of problems in \mainds that for at least one of their corresponding submission pairs $\langle P, Q \rangle$, T generates a DET. A tool that generates DETs for more problems is effective on a more diverse set of programs, indicating strong external validity.

% Finally, we assess the performance of T by measuring how many hours it takes to finish the experiment on \mainds.

This experiment is conducted on a machine with 128 AMD EPYC 7742 64-Core processors, running at 2.2GHz and having eight 16GB DDR4 RAMs with 3.3GHz speed.

\subsection{Protocol for RQ2 (ablation study)}
\label{sec:rq2_protocol}

To answer RQ2, we run \toolname on \mainds and count $\#Success\_Pairs$, explained in \autoref{sec:rq1_protocol}, with various settings as follows:
\begin{itemize}
    \item To measure the effect of \emph{descriptions}, we run \toolname without providing the program descriptions in the prompt. % batie-it-10-samp-10-temp-1
    \item To measure the effect of \emph{randomness} in LLM's responses, we run \toolname with temperature=0. This leads to more deterministic responses, meaning we generate only one sample at each iteration for this setting. % badtie-it-10-samp-1-temp0
    \item To measure the effect of \emph{LLM}'s accuracy and check if \toolname is well designed to work with various LLMs, we also run \toolname using three other LLMs. \modifiedtwo{First, we use two open-source LLMs from two different enterprises: \texttt{CodeLlama-instruct-7b} by Meta and \texttt{Codestral-2501} by Mistral AI. Codestral is one of the most recent LLMs as of April 2025; it is a lightweight, low-latency model designed for code and reasoning tasks. This makes Codestral an ideal LLM for evaluating how far we can push the effectiveness of \toolname using affordable models.} To keep the resource demand of this experiment manageable, we run this experiment with the minimum temperature and generate one sample at each iteration. In addition to the open-source models, we also run \toolname with \texttt{gpt-4o-2024-05-13}, as one of the most powerful existing LLMs. We use GPT-4o with two configurations: \modifiedtwo{(1) the same configuration as for CodeLlama and Codestral, to compare all models with each other,} and (2) the default configuration used by \toolname, to assess \toolname's effectiveness when using the most advanced existing LLM. % codellama & gpt-4o
    \item To measure the effect of \emph{example test}, we run \toolname without providing the example test and its corresponding output and execution difference in the \initprompt and \iterprompt prompts. % badt-it-10-samp-10-temp-1
    \item To measure the effect of \emph{execution data}, we run \toolname without providing the output and execution difference in the \initprompt and \iterprompt prompts.
\end{itemize}

\subsection{Protocol for RQ3 (impact of P and Q characteristics)}
\label{sec:rq3_protocol}
\modifiedtwo{To answer RQ3, we assess the relation between the effectiveness of \toolname and four different characteristics of the input to the tool: the total number of tokens in P and Q, the cyclomatic complexity of Q, the Levenshtein distance between P and Q, and the number of tokens in the longest test for program pair in \mainds.} Note that \mainds contains multiple tests for each problem in \mainds. We consider the longest test in this experiment as it suggests how complex the input to the program pair can be.

Consider the number of tokens of P and Q as an example. We assess its relation with the effectiveness of \toolname as follows. First, for each $\langle P, Q \rangle$ pair in \mainds, we compute the number of tokens in P and Q with the gpt-3.5 tokenizer. Next, we split all pairs in \mainds into 10 subsets with the same sizes, S1, S2, ..., S$i$, ..., S10, where $i$ is the index of the $i$-th subset S$i$. S1 contains 10\% of pairs with the fewest tokens, while S10 contains 10\% of pairs with the most tokens. Next, we compute the percentage of pairs in each subset for which \toolname generates a DET, and call it $SSuccess$. Finally, we run a Spearman statistical test with p-value=0.05 to check if there is a statistically significant correlation between the subset's index and its $SSuccess$.

\modifiedtwo{To evaluate the relation between DET generation effectiveness and the three other input characteristics (cyclomatic complexity of Q, distance between P and Q, and number of tokens in the example test), we follow a similar strategy.} The only difference is the metric according to which we sort the $\langle P, Q \rangle$ pairs and split the data into ten subsets.
\modifiedtwo{A significant correlation between $SSuccess$ and each of the four considered metrics provides valuable information regarding where we should expect \toolname to work in practice.} It can also inform us of the fundamental reasons why LLMs can generate DETs in some cases and fail to do so in other cases.

\section{Experimental Results}
\label{sec:results}

\subsection{Results for RQ1 (effectiveness)}
\label{sec:rq1_results}

% det_methods: pynguin-c4b=292, 
\begin{table*}[t]
\centering
\tiny
\caption{Effectiveness of \toolname compared with related work. QuixBugs dataset contains \qbsize pairs of programs, and \mainds contains \maindssize pairs of programs for \maindsproblems problems. \#Success\_pair is the number of pairs for which a DET is generated, \#TT is the number of all generated tests, and \#DET\_Pr is the number of problems a DET is generated for at least one of the corresponding pairs.}
\label{tab:effectiveness}
\begin{tabular}{@{}l | l l l | l l r @{}}
    \toprule
    & \multicolumn{3}{c}{QuixBugs} & \multicolumn{3}{c}{\mainds} \\
    Tool & \#Success\_Pair & \#TT & \#DET\_Pr & \#Success\_Pair & \#TT & \#DET\_Pr \\
    \midrule
    \pynguin & 50.0\% (16/\qbsize) & 1,176 & 50.0\% (16/\qbsize) & 4.9\% (76/\maindssize) & 27,487 & 18.5\% (41/221) \\
    \dpp & 50.0\% (16/32) & 3,200 & 50.0\% (16/32) & 37.3\% (573/\maindssize) & 153,500 & 70.5\% (156/221) \\
    \hline
    \toolname (iter 1) & 93.6\% (30/32) & 320 & 93.6\% (30/32) & 58.9\% (905/\maindssize) & 15,350 & 87.7\% (194/221) \\
    \toolname (iter 4) & 100.0\% (32/32) & 340 & 100.0\% (32/32) & 76.4\% (1,173/\maindssize) & 30,260 & 90.9\% (201/221) \\
    \toolname (iter 7) & 100.0\% (32/32) & 340 & 100.0\% (32/32) & 80.3\% (1,234/\maindssize) & 40,240 & 92.7\% (205/221) \\
    \textbf{\toolname} & \textbf{100.0\% (32/32)} & \textbf{340} & \textbf{100.0\% (32/32)} & \textbf{81.7\% (1,255/\maindssize)} & \textbf{48,930} & \textbf{94.1\% (208/221)} \\
    \bottomrule
\end{tabular}
\end{table*}

\begin{figure}
\begin{center}
\includegraphics[width=0.6\textwidth]{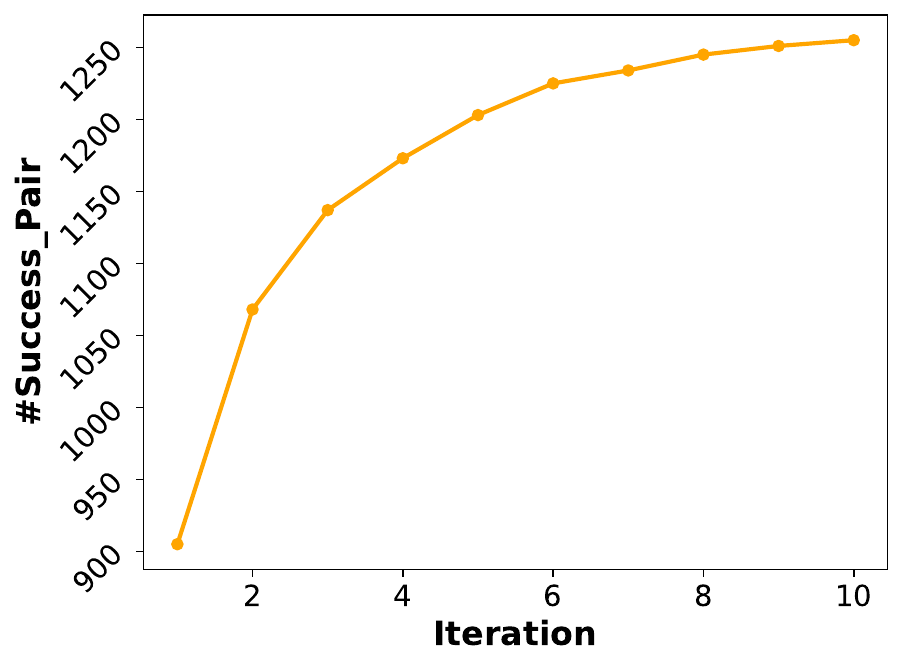}
\caption{Effectiveness over iteration. \toolname drives the LLM towards finding difference-exposing tests.}
\label{fig:iter-success}
\end{center}
\end{figure}

% ./generated_tests/819/8865/temp_test_case_9.py
\begin{lstlisting}[float=tb, style=diff, caption={A pair of programs for which only \toolname is able to generates a complex DET, after several iterations.}, label=lst:outperformance_ex]
%{\color{blue} \textbf{\textbf{The diff between P \& Q}:}}%
s = input()
s1 = [i for i in range(len(s)) if ((s[i]=='A') or (s[i]=='E') or (s[i]=='I') or (s[i]=='O') or (s[i]=='U') or (s[i]=='Y'))]
d = 0
for i in range((len(s1) - 1)):
  if ((s1[(i + 1)] - s1[i]) > d):
    d = (s1[(i + 1)] - s1[i])
if ((d == 0) and (len(s1) != 0)):
  print(max((s1[0] + 1), (len(s) - s1[(- 1)])))
elif (len(s1) == 0):
  print((len(s) + 1))
else:
%\RHilight%  d = max(d, (len(s) - s1[(- 1)]))
%\GHilight%  d = max(d, (len(s) - s1[(- 1)]), (s1[0] + 1))
  print(d)
%\hrule%
%{\color{blue} \textbf{\textbf{DET generated by \toolname}:}}%
Input:
    "BYAIAUOIEOAA"
Output:
    P: 1
    Q: 2
\end{lstlisting}

% mokav misses 41 cases that dpp covers, and 6 that pynguin covers. dpp misses 24 of pynguin dets

\autoref{tab:effectiveness} compares the effectiveness of \toolname on QuixBugs and C4DET with two baselines (Pyguin and \dpp). For each row, the first column is the name of the tool. The table presents \toolname's effectiveness with 1, 4, and 7 iterations, and its default setting with 10 iterations for the last row. For each dataset, we measure three metrics: the number of program pairs with DETs (``\#Success\_Pair''), the total number of generated tests (``\#TT''), and the number of problems that the tool generates a DET for at least one of its corresponding program pairs (``\#DET\_Pr''), respectively.

% The last three columns of \autoref{tab:effectiveness} show the effectiveness of \toolname on QuixBugs, in terms of ``\#Success\_Pair'', ``\#TT'', and ``DET\_R'', respectively.

%The second column of \autoref{tab:effectiveness} shows that \toolname outperforms the baselines on the QuixBugs dataset. We note that \toolname generates a DET for 100\% (40/40) of pairs of programs in QuixBugs at its first iteration. This is significantly higher than 50\% (20/40) success rate by \pynguin and also higher than 92\% (37/40) effectiveness of \dpp. A 100\% success by \toolname (iter 1) on QuixBugs indicates that QuixBugs' program pairs are too easy to challenge \toolname's DET generation ability. This necessitate using a larger and more complex dataset, like \mainds, for evaluating test generation with LLMs.

According to \autoref{tab:effectiveness}, \toolname generates a DET for 58.9\% (905/\maindssize) of \mainds pairs at its first iteration and for 81.7\% (1,535/\maindssize) of pairs at its last iteration, which is the highest performance ever reported. First, this shows that the iterative approach significantly improves the effectiveness of \toolname. This is also illustrated in \autoref{fig:iter-success}. \modifiedtwo{Second, this result shows that even at its initial attempt, \toolname outperforms both strong baselines \pynguin and \dpp: \ toolname's initial attempt generates a DET for 58.9\% of pairs, while \pynguin and \dpp generate DETs for 4.9\% and 37.3\% of program pairs, respectively. While the iteration phase significantly improves \toolname effectiveness, \toolname achieves strong DET generation results even without paying the cost of iterative LLM invocations.} Our manual investigation reveals that \toolname outperforms \pynguin and \dpp mainly because it better understands the type of arguments the programs take. This confirms the importance of using an example test in the prompt to hint to the LLM regarding the type and structure of inputs.

We notice that \toolname, due to its iterative and execution-driven feedback, better understands the semantic differences between P and Q. \autoref{lst:outperformance_ex} shows an example where \toolname generates a DET at its final iteration, while \pynguin, and \dpp fail to do so. P and Q are supposed to compute the distance between certain characters in a string and the beginning and end of a given string, and output the longest distance. P wrongly ignores the distance between the beginning of the string and the first appearance of the considered characters (line 13). \toolname successfully generates ``BYAIAUOIEOAA'' as a DET for this pair. The DET is only generated at the last iteration of \toolname. This indicates that the iterative and execution-driven approach of \toolname enables it to better understand P and Q's semantics compared to \dpp and \pynguin.

The next column (``\#TT'') shows the total number of tests generated by each tool. \dpp generates 153,500 tests for \maindssize pairs in \mainds, because it asks for a fixed number of 100 tests for each pair. \toolname generates batches of 10 tests at each iteration only if the previous iterations fail to yield a DET. Consequently, \toolname generates only 15,350 tests in its first iteration, and this number gradually increases to 48,930 at its 10th iteration. This shows that \toolname's iterative approach also contributes to its cost efficiency by generating new tests step-by-step and only when needed.

%The $DET\_R$ also reaffirms the cost efficiency of \toolname. Note that $DET\_R$ decreases as \toolname runs for more iterations, from 30.1\% to 10.3\%. $DET\_R$ gradually declines because when \toolname fails to generate a DET for several iterations, it becomes more and more unlikely for it to generate a DET in later iterations. We conclude that an iterative approach for DET generation makes the tool more effective; however, the number of iterations should not be excessively high.

The ``\#DET\_Pr'' column of \autoref{tab:effectiveness} presents the number of problems for which the tool generates a DET for at least one of their corresponding program pairs. Recall that the program pairs in \mainds are submissions to answer online competition problems. A problem may have many corresponding program pairs in \mainds (see \autoref{sec:dataset}). In QuixBugs, there is exactly one pair per problem; therefore, \#Success\_Pair and \#DET\_Pr are the same for QuixBugs. \toolname outperforms both baselines in terms of $\#DET\_Pr$ by generating a DET for 94.1\% (208/\maindsproblems) of the problems. This shows that \toolname's effectiveness is not limited to small and specific types of programs; it can generate DETs for a diverse set of programs.

All the major takeaways from the results on \mainds also hold for QuixBugs. Most importantly, \toolname outperforms both \pynguin and \dpp even at its first iteration. \toolname generates a DET for 100\% (32/32) of the program pairs in QuixBugs. This demonstrates the perfect effectiveness of \toolname on this widely-used dataset. \modified{We further analyze our experimental results to understand the reasons behind this perfect effectiveness. We first notice that the effectiveness of \toolname cannot be attributed to data leakage in the LLM training dataset, as \dpp also uses the same LLM for test generation and fails in 50\% (16/32) of the cases. We discuss the threats of data leakage in \autoref{sec:threats}. Our analysis reveals that QuixBugs DETs are close to the example tests in the dataset. Therefore, adding the example test to \toolname prompts significantly helps the LLM in generating DETs. This finding  explains the difference between the effectiveness of \dpp and \toolname, which lies in the provided example test.}

% We also measure the time each tool spends to complete DET generation on \mainds. \pynguin, \dpp, and \toolname take 43, 1.5, and 7 hours to finish the experiment, respectively. We note that \dpp is faster than \toolname for two main reasons. First, \toolname at first generates a description for each version P and Q. This means \toolname invokes to the LLM two times more than \dpp at each iteration. Second, \toolname has an iterative approach in which it calls the LLM up to ten times to generate a DET. In contrast, \dpp calls the LLM only once. Note that \toolname spends seven hours on \maindssize pair of programs, or 27 seconds per pair on average. This suggests that \toolname is fast enough to be integrated into the development process.

\begin{mdframed}\noindent
    \textbf{Answer to RQ1: \rqone} \\
    We evaluate \toolname, \pynguin, and \dpp for generating difference exposing tests on \qbsize program pairs from QuixBugs and \maindssize program pairs from \mainds. The results clearly show that \toolname outperforms the state of the art on both datasets. On QuixBugs, \toolname achieves a perfect score by generating DETs for 100\% (31/\qbsize) of cases. On \mainds, \toolname produces DETs for 81.7\% (1,255/\maindssize) of the program pairs. Our results also demonstrate that \toolname's effectiveness improves over iterations, starting at 58.9\% and improving up to 81.7\%. 
    To our knowledge, \toolname is the best performing approach and tool for generating DETs.
\end{mdframed}

\subsection{Results for RQ2 (ablation study)}
\label{sec:rq2_results}

% badtie-it-10-sam-10-temp-1 => 1580
% badti-it-10-sam-10-temp-1 => 1547
% aat-it-1-sam-100-temp-1 => 757
% batie-it-10-sam-10-temp-1 => 1560
% badtie-it -10-sam-1-temp-0 => 1095
% gpt-4-badtie-it-10-sam-1-temp-0 => 1,459
% badt-it-10-sam-10-temp-1 => 524

\autoref{tab:rq2_effectiveness} presents the results of our ablation study. It compares the effectiveness of different \toolname configurations in terms of their \#Success\_Pairs. \modifiedtwo{Nine configurations, C1-C9, are considered, with C1 being the default setup of \toolname.} C2, C3, and C4 represent configurations where the description, the example test, and the execution data are excluded from the prompt, respectively. Note that in C3, as the example test is excluded, the execution data that depends on the presence of an example test is excluded as well. In C5, temperature=0 is used and only the top one sample is requested from the model at each iteration. \modifiedtwo{C6, C7 and C8 are similar to C5, but use different LLMs, CodeLlama, Codestral, and GPT-4o, respectively.} Finally, C9 has the same configuration as C1, but uses GPT-4o as the LLM. The last column of \autoref{tab:rq2_effectiveness} shows the \#Success\_Pairs, the number of pairs in \mainds for which each configuration generates a DET.

C2, C3, and C4 show that the presence of each component of the prompt contributes to the effectiveness of \toolname. C3 results show that removing the example test and its corresponding execution data has the highest impact on effectiveness by decreasing it from 81.7\% to 26.3\%. Our manual investigation reveals the main reason for this decline: by removing the example test, the LLM cannot recognize the type and structure of the input to be generated for the program pairs. For example, the example test hints to the LLM if the inputs should be strings or integers, what should be the content of string inputs, how many inputs are required, \etc In the absence of the one-shot example test, all this information is missing from the prompt and this negatively affects the effectiveness. Overall, the example test has the most significant impact.

The effectiveness of C5 is 51.7\%, versus 81.7\% for C1. This shows the large impact of using a high temperature and sample size. By using a high temperature and sample size, we can better exploit the LLM's capability to explore the input space of program pairs.

\begin{table*}[t]
\centering
\tiny
\caption{\modifiedtwo{Ablation study about the effectiveness of different \toolname components.}}
\label{tab:rq2_effectiveness}
\begin{tabular}{@{}l | l l l l c c c r @{}}
    \toprule
    Config ID & LLM & Iter. & Samp. & Temp. & Desc. & Ex\_Test & Exec\_Data & \#Success\_Pairs \\
    \midrule
    \textbf{C1 (\toolname)} & \textbf{GPT-3.5} & \textbf{10} & \textbf{10} & \textbf{1} & \textbf{\cmark} & \textbf{\cmark} & \textbf{\cmark} & \textbf{81.7\% (1,255/\maindssize)} \\
    \hline
    C2 & GPT-3.5 & 10 & 10 & 1 & \xmark & \cmark & \cmark & 80.4\% (1,235/\maindssize) \\
    C3 & GPT-3.5 & 10 & 10 & 1 & \cmark & \xmark & \xmark & 26.3\% (404/\maindssize) \\
    C4 & GPT-3.5 & 10 & 10 & 1 & \cmark & \cmark & \xmark & 79.4\% (1,220/\maindssize) \\
    \hline
    C5 & GPT-3.5 & 10 & 1 & 0 & \cmark & \cmark & \cmark & 51.7\% (795/\maindssize) \\
    C6 & CodeLlama & 10 & 1 & 0 & \cmark & \cmark & \cmark & 11.4\% (175/\maindssize) \\
    \modifiedtwo{C7} & \modifiedtwo{Codestral} & \modifiedtwo{10} & \modifiedtwo{1} & \modifiedtwo{0} & \modifiedtwo{\cmark} & \modifiedtwo{\cmark} & \modifiedtwo{\cmark} & \modifiedtwo{64.4\% (989/\maindssize)} \\
    C8 & GPT-4o & 10 & 1 & 0 & \cmark & \cmark & \cmark & 73.9\% (1,135/\maindssize) \\
    C9 & GPT-4o & 10 & 10 & 1 & \cmark & \cmark & \cmark & 85.5\% (1,313/\maindssize) \\
    \bottomrule
\end{tabular}
\end{table*}

%def program(n: int, k: int, a1: str) -> list[str]:
\begin{lstlisting}[float=tb, style=diff, caption={Code diff between a pair of programs for problem 226B. A DET for this pair of programs should be an input string that contains exactly 50 characters. GPT-4o is the only model that generates a DET for this program pair.},  label=lst:gpt4-ex]
%{\color{blue} \textbf{\textbf{The diff between P and Q}:}}%
# P and Q get two integers (n and k) and a string (a1) as input and return a list of strings as output.
...
%\RHilight%  - if ((1 <= n <= 50) and (1 <= k <= 50)):
%\GHilight%  + if ((1 <= n < 50) and (1 <= k <= 50)):
...
%\hrule%
%{\color{blue} \textbf{\textbf{The DET generated by \toolname:}}%
{n=50, k=1, a1='BBBBBBBBBBBBBBBBBBBBBBBBBBBBBBBBBBBBBBBBBBBBBBBBBG' }
%\hrule%
%{\color{blue} \textbf{\textbf{The output produced by P and Q on the generated DET:}}%
P output: []
Q output: ['BBBBBBBBBBBBBBBBBBBBBBBBBBBBBBBBBBBBBBBBBBBBBBBBGB']
\end{lstlisting}

% changing the MM%
\modifiedtwo{C5, C6, C7, and C8 use different LLMs. The table shows that the model in use impacts effectiveness. 
CodeLlama underperforms our default GPT-3.5 with an effectiveness of only 11.4\% (C6 $\ll$ C5). In contrast, Codestral outperforms GPT-3.5 (C7 $\gg$ C5) by a large margin, succeeding on 64.4\% (989/1,535) of the program pairs. We note that the total cost of DET generation with Codestral for \maindssize pairs in our dataset was 6.45 USD. This indicates the practicality of \toolname, as it achieves strong effectiveness by leveraging affordable open-source models that are optimized for coding tasks. Also, as expected, the highly strong and quite expensive GPT-4o surpasses GPT-3.5 (C8 $\gg$ C5), CodeLlama (C8 $\gg$ C6), and Codestral (C8 $\gg$ C7). This demonstrates \toolname's potential to harness the power of frontier models for DET generation.}

\modifiedthree{Given the low cost of open-source models, such as Codestral, we can increase the sample size at each iteration while keeping the test generation cost reasonable. As our comparison between C1 and C5 shows, generating more samples can improve the effectiveness of \toolname. Therefore, we expect to achieve greater effectiveness by increasing the Codestral sample size, even beyond the \toolname default configuration of $sample\_size=10$. This technique can be evaluated in the future to see if we can replace expensive closed-source models with affordable transparent open-source ones without losing effectiveness.}

\modifiedtwo{Finally, C9 shows the results for the last experiment, where we run \toolname with default settings similar to C1 using GPT-4o, the best model in our experiments. 
The results show that effectiveness is improved to 85.5\%, the highest among all configurations. This improvement compared to C1 shows that using a stronger LLM that better understands the semantics of programs directly impacts effectiveness.} This significant improvement reaffirms the soundness of the default setting used in \toolname: a high sample size (10) and high temperature (10) lead to generating a diverse set of useful test cases.

We now discuss an interesting case in \autoref{lst:gpt4-ex}, which shows the code diff between a pair of programs, and input and output. 
The pair comes from author ``60,724'' for problem 226B on Codeforces.
\modified{Among all the configs in \autoref{tab:rq2_effectiveness}, only C8, which uses GPT-4.o with temperature=0 and ten iterations of execution-based feedback, succeeds in generating a valid DET. Our analysis also shows that \dpp and \pynguin fail to generate a valid DET in this case. The valid DET generated by C8 contains exactly 50 characters as input (``BBBBBBBBBBBBBBBBBBBBBBBBBBBBBBBBBBBBBBBBBB\\BBBBBBBG'').
This example demonstrates two facts. First, the increasingly more powerful LLMs are valuable for generating DETs, as they better understand the code semantics. Secondly, this example shows that in out-of-distribution cases the execution-based feedback is needed on best LLM solutions for guiding the LLM toward DET generation.}

On the negative side of results, there are 11.5\% (178/\maindssize) of pairs that no config generates a DET for. \autoref{lst:no-det} is one of these pairs. Per our manual analysis, in most of these pairs, similar to \autoref{lst:no-det}, the LLM needs to perform precise mathematical computations to generate a DET. This reveals a core limitation of LLMs that affects their performance on code related tasks.

\begin{mdframed}\noindent
    \textbf{Answer to RQ2: \rqtwo} \\
    \modifiedtwo{Our comparison of nine different configurations of \toolname shows that using \toolname's optimized setup with a frontier model (GPT-4o) surpasses all other configurations with a maximum effectiveness of 85.5\% (1,313/\maindssize).} Among components in \toolname's approach, the example test given to the LLM plays the most important role. Using a high temperature and sample size effectively leverages the LLM's ability to explore various parts of the input space.
\end{mdframed}

\subsection{Results for RQ3 (impact of P and Q characteristics)}
\label{sec:rq3_results}

%def program(n: int, k: int, a1: str) -> list[str]:
\begin{lstlisting}[float=tb, style=diff, caption={A pair for which no config generates a DET.}, label=lst:no-det]
%{\color{blue} \textbf{\textbf{The diff between P and Q}:}}%
n, k = map(int, input().split())
%\RHilight%s = int(n / k)
%\GHilight%s = int(n // k)
if s %\%% 2 != 0:
...
\end{lstlisting}

\autoref{tab:rq3-correlation} shows the results of the RQ3 experiment. In this table, S1-S10 represent the subsets of program pairs after split according to each considered program pair characteristics. \modifiedtwo{The table has four main vertical sections representing each of those four characteristics: ``SRC\_Tok'', the number of tokens in a program pair; ``Cyclo\_Comp'', the cyclomatic complexity of Q, which represents the correct version of the program pair in \mainds; ``Test\_Tok'', the number of tokens in the longest test in \mainds for a program pair; and ``Lev\_Diff'', the Levenshtein distance between a pair of programs.} For each characteristic, the table shows the median value for each subset (``Med'') and the ratio of pairs in each subset for which \toolname successfully generates a DET (``SSuccess''). For example, consider S1. The median number of tokens in the 10\% shortest pairs is 74. \toolname generates a DET for 96\% of these program pairs. \modifiedtwo{The median cyclomatic complexity of the 10\% least complex programs is 1, and \toolname generates a DET for 91\% of these program pairs.} The median number of tokens in the test for 10\% of program pairs that have the shortest test is 1. \toolname generates a DET for 89\% of these program pairs. Finally, the median Levenshtein distance for the 10\% of program pairs with the shortest distance is 0.44. For 92\% of these program pairs, \toolname successfully generates a DET.

% spearsman interpretation https://www.statstutor.ac.uk/resources/uploaded/spearmans.pdf
\begin{table}[t]
\centering
\footnotesize
\caption{\modified{The correlation between \toolname's effectiveness and P \& Q characteristics.}}
\label{tab:rq3-correlation}
\begin{tabular}{@{}l | l r | l r | l r | l r @{}}
    \toprule
    & \multicolumn{2}{c}{SRC\_Tok} & \multicolumn{2}{c}{\modifiedtwo{Cyclo\_Comp}} & \multicolumn{2}{c}{Test\_Tok} & \multicolumn{2}{c}{Lev\_Diff}  \\
    & Med & SSuccess & \modifiedtwo{Med} & \modifiedtwo{SSuccess} & Med & SSuccess & Med & SSuccess \\
    \midrule
    S1 & 74 & 0.96 & \modifiedtwo{1} & \modifiedtwo{0.91} & 1 & 0.89 & 0.44 & 0.92 \\
    S2 & 115 & 0.88 & \modifiedtwo{2} & \modifiedtwo{0.86} & 2 & 0.84 & 0.68 & 0.90 \\
    S3 & 153 & 0.84 & \modifiedtwo{3} & \modifiedtwo{0.80} & 3 & 0.79 & 0.80 & 0.86 \\
    S4 & 187 & 0.86 & \modifiedtwo{4} & \modifiedtwo{0.80} & 4 & 0.77 & 0.87 & 0.76 \\
    S5 & 220 & 0.77 & \modifiedtwo{4} & \modifiedtwo{0.80} & 6 & 0.77 & 0.91 & 0.76 \\
    S6 & 263 & 0.79 & \modifiedtwo{5} & \modifiedtwo{0.79} & 9 & 0.83 & 0.94 & 0.79 \\
    S7 & 312 & 0.77 & \modifiedtwo{6} & \modifiedtwo{0.83} & 13 & 0.78 & 0.96 & 0.77 \\
    S8 & 384 & 0.81 & \modifiedtwo{7} & \modifiedtwo{0.78} & 21 & 0.68 & 0.99 & 0.80 \\
    S9 & 496 & 0.75 & \modifiedtwo{9} & \modifiedtwo{0.81} & 55 & 0.84 & 0.99 & 0.79 \\
    S10 & 809.5 & 0.72 & \modifiedtwo{14} & \modifiedtwo{0.72} & 82 & 0.92 & 0.99 & 0.78 \\
    \midrule
    Cor & \multicolumn{2}{c}{-0.89} & \multicolumn{2}{c}{\modifiedtwo{-0.63}} & \multicolumn{2}{c}{-0.01} & \multicolumn{2}{c}{-0.45} \\
    p-val(<0.05) & \multicolumn{2}{c}{0.0004 (\cmark)} & \multicolumn{2}{c}{\modifiedtwo{0.04 (\cmark)}} & \multicolumn{2}{c}{0.97 (\xmark)} & \multicolumn{2}{c}{0.18 (\xmark)} \\
    \bottomrule
\end{tabular}
\end{table}

\modifiedtwo{The last two rows of \autoref{tab:rq3-correlation} show the correlation between the subset index and each of the considered characteristics. There is a strong negative correlation of -0.89 between the number of tokens in the source code and the success rate of \toolname. Similarly, there is a strong negative correlation of -0.63 between the cyclomatic complexity of the program and the success rate of \toolname. Both of these correlations are statistically significant with a p-value of 0.0004 and 0.04, respectively. These statistically significant negative correlations highlight that \toolname is less likely to generate a valid DET for more syntactically complex program pairs.
In sum, \autoref{tab:rq3-correlation} shows that the length and complexity of program pairs are the only characteristics with a statistically significant correlation with the success rate of \toolname. This indicates that to apply \toolname on lengthy and complex multi-function programs, we would have to employ powerful LLMs with large context windows, and delicately provide them with runtime hints regarding program semantics.}

The absence of a significant correlation between the length of the longest test in \mainds for a program pair and \toolname's success rate demonstrates the following: the size of the inputs for a program pair does not directly tell how challenging it is to find the corner cases that expose differences. 
We observe no significant correlation between the program pair's distance and \toolname's success rate. At first glance, it is easier to expose the difference between two programs if they are textually distant; therefore, we expect to see a positive and significant correlation between \toolname's success rate and the distance between pairs. On the other hand, we notice that in \mainds, program pairs close to each other tend to be smaller and easier to understand. This means we expect the LLM to better understand program pairs with small distances and thus generate a valid DET for them. Consequently, the distance between program pairs impacts \toolname's success rate from two opposite directions. This leads to the absence of a significant correlation between \toolname's effectiveness and distance between P\&Q.

\begin{mdframed}\noindent
    \textbf{Answer to RQ3: \rqthree} \\
    \modifiedtwo{Our experiments show a significant correlation between the length and cyclomatic complexity of program pairs and \toolname's success rate in generating a difference exposing test. There is strong statistical evidence that LLMs better capture the semantics of syntactically simple program pairs to find corner cases that expose their differences.} We do not find a significant correlation between \toolname's effectiveness and the distance between program pairs.
\end{mdframed}

\section{Discussion}
\label{sec:threats}

\subsection{Threats to Construct Validity}
The main threat to the construct validity of this study is data leakage in LLMs~\cite{al2024traces}. As LLMs are trained on a large corpus of open-source data, they may have some program pairs of our datasets in their training data. This may threaten the generalizability of our experimental conclusions. Despite this concern, we note that open-source repositories usually do not put difference-exposing tests next to programs. This means there is no direct connection between programs and DETs in LLMs' training data. Consequently, the data leakage problem is less significant in DET generation, the goal of \toolname, compared to more traditional tasks, such as code generation~\cite{du2024evaluating} and program repair~\cite{jiang2023impact}. Also, to further alleviate the data leakage concern, we assess the effectiveness of \toolname on two different datasets, \ie, QuixBugs~\cite{lin2017quixbugs} and \mainds.

\subsection{Threats to Internal Validity}
\toolname uses a high temperature for the LLM in its default configuration. This leads to randomness in its responses, a threat to the internal validity of our experimental results. To address this threat, we repeat our main experiment of generating DETs for program pairs in \mainds five times. The results show that in all five instances, the effectiveness of \toolname is between 80.1\% and 81.7\%. This indicates that the main conclusions of our study are robust and only slightly affected by the randomness of LLMs.

\subsection{\modifiedtwo{Threats to External Validity}}
\modifiedtwo{In this study, we evaluated \toolname on a dataset of program pairs with three constraints: (1) program length, (2) program language, and (3) program input types. These constraints might limit the generalizability of our results. Now we discuss each constraint, its implications, and potential solutions.}

\subsubsection{\modifiedtwo{Program Length}}
\modifiedtwo{As explained in \autoref{sec:dataset}, the program pairs in our dataset contain at most 5,000 tokens, 2,500 for each version. We consider this limit for our dataset to avoid exceeding the context size of widely-used LLMs, such as GPT-3.5 with a context size of 16,000 tokens. This decision might threaten the generalizability of our results to real-world programs. We believe it is not a blocker for two reasons. First, previous studies show that a large number of important code changes, such as bug fixes, occur at the single-function level~\cite{karampatsis2020often}, which are often small. This indicates the practical usefulness of generating DETs for program pairs with single-function diffs, which mostly have fewer than 5,000 tokens. Second, recent work highlights the possibility of using LLMs on coding tasks with a divide-and-conquer strategy~\cite{chen2024divideandconquer}. 
This suggests that one could generate DETs for large programs by first splitting them into smaller units, trying to find a DET for one of the smaller units, and then assembling the results to obtain a DET for the full program. Using \toolname as a core component in a divide-and-conquer strategy for generating DETs for large programs is a promising direction for future work.}

\subsubsection{\modifiedtwo{Program Language}}
\modifiedtwo{In this study, we evaluate our approach on a dataset of Python programs. We focus on Python programs as the closest related work~\cite{li2023nuances,liu2024llm} and our main baseline (DPP~\cite{li2023nuances}) also evaluate their tools on Python programs. To have a quantitative and fair comparison, we have to consider the same language. This poses a threat to the generalizability of our results to other programming languages. We note that recent studies have shown that LLM-based tools are capable of multi-language test generation~\cite{pan2025aster}. This indicates the potential of extending \toolname to generate DETs for other programming languages. \modifiedthree{We believe that the main challenge for such an extension is collecting execution differences for other languages. There are existing tools for execution differencing in other programming languages, such as Java \cite{etemadi2023augmenting}, but we still need to adapt these tools and integrate them into the \toolname workflow.} We consider this an opportunity for future work.}

\subsubsection{\modifiedtwo{Program Input Types}}

\modifiedtwo{The current version of \toolname is focused on generating inputs of primitive types (i.e., integers, floats, characters, and strings) and lists of such primitive data. This is a threat to the applicability of our approach to real-world programs, as many real-world projects have inputs with complex data types. Nevertheless, we believe that our proposed approach has significant real-world applications for two reasons. First, among the program pairs taken from the QuixBugs~\cite{lin2017quixbugs} dataset (see \autoref{sec:dataset}), four program pairs implement graph algorithms, such as the breadth first search. To generate DETs for these pairs, \toolname must produce inputs of the complex \texttt{Node} type. For this purpose, \toolname had to be modified to support generating \texttt{Node} inputs. One of the authors performed this modification with less than an hour of engineering effort. This shows that \toolname provides a flexible framework that can be quickly extended to generate DETs for real-world programs that use complex data types. Second, we note that recent studies demonstrate the strength of LLMs in generating tests with complex data types~\cite{gu2025llm}. This is promising for \toolname to produce DETs for real-world projects with non-primitive inputs. We consider such an improvement as one of the main paths for future work.}

\subsection{\modifiedtwo{Safety of LLM-generated Tests}}
\modifiedtwo{As the adoption of LLMs for coding is growing, researchers have raised concerns about the safety of LLM-generated code~\cite{tihanyi2025secure}. They note that code snippets generated by LLMs can contain vulnerabilities, such as incorrect exception handling or well-known CWEs~\cite{ramirez2024state}. Such safety concerns also apply to tools like \toolname that employ LLMs for test generation and execute all outputs. For example, the execution of an LLM-generated code modifying file permissions could compromise the tester system. Since \toolname generated tests just call the program pairs with various inputs and lack a large test body, they are less prone to introducing such security problems. \modifiedthree{However, we still need to cautiously run LLM-generated tests to avoid potential issues, using safeguards such as rule-based validation of generated tests~\cite{panichella2024vulnerabilities} or running tests in isolated sandbox environments~\cite{dou2024multi}, like container-based sandboxes \cite{khalimov2019container}}.}

\section{Related Work}
\label{sec:related_work}

%\subsection{Test Generation with LLMs}
%\label{sec:test-with-llms}

\subsection{Test Generation with LLMs}
LLMs are being applied on many software engineering tasks~\cite{fan2023large}, and test generation is one of the major targets~\cite{wang2024software,siddiq2023exploring,xie2023chatunitest,tang2024chatgpt,lemieux2023codamosa,tufano2020unit,altmayer2024coverup,chen2022codet,takerngsaksiri4736450pytester,plein2023automatic,lahiri2022interactive,vikram2023can,liu2024make,nie2023learning,ryan2024code,dakhel2024effective,pizzorno2024coverup,alshahwan2024automated,steenhoek2023reinforcement}. A test consists of multiple parts: an input to a program under test (PUT), a test setup that moves the PUT into a specific state, and an oracle that asserts the PUT correctly produces the expected outputs. LLMs are used to generate all these parts of tests.

Deng~\etal~\cite{deng2023large} introduce TitanFuzz. This tool employs LLMs to generate human-like code that tests the API of deep learning (DL) libraries. TitanFuzz uses LLMs to generate and mutate input DL programs for fuzzing. Xia~\etal~\cite{xia2024fuzz4all} propose Fuzz4All, an LLM-based fuzzer that generates input test programs for compilers. Fuzz4All iteratively prompts the LLM to create new test inputs. The authors show that Fuzz4All effectively generates test inputs for six different programming languages. While TitanFuzz and Fuzz4All generate input programs for a single system, \toolname generates test inputs that differentiate two given programs.

LLMs have also been used for oracle generation~\cite{watson2020learning,tufano2022generating,hossain2024togll,li2024large}. Dinella~\etal propose TOGA~\cite{dinella2022toga}. TOGA uses a fine-tuned version of CodeBERT~\cite{feng2020codebert} to generate assertion candidates based on the PUT and test prefix and rank these candidate assertions. Liu~\etal~\cite{liu2023towards} show limitations of TOGA: its unrealistic assumption of the availability of a correct focal method, its ignorance of important evaluation metrics, such as precision, and the lack of a straightforward baseline. Zhang~\etal~\cite{zhang2024generating} generate tests in which the oracle expects an exception. In contrast to these works, \toolname focuses on input generation instead of oracle generation.

% Nie~\etal~\cite{nie2023learning} introduce TECO whose goal it to complete an already existing test method. TECO extracts six kinds of semantic data, including execution result of previous test statements and adds this data to the existing test and code context. Based on all this information, the LLM is asked to predict the next statement in the test method. Ryan~\etal~\cite{ryan2024code} also add execution path information to the prompt to make the LLM generate tests that cover more paths in the program.

%Dakhel~\etal~\cite{dakhel2024effective} introduce MuTAP, an LLM-based tool that improves the effectiveness of generated tests. To improve test effectiveness, MuTAP generates new tests that kill survived mutants. In contrast with MuTAP, \toolname is not focused on improving existing tests for a single program, instead it generate tests that expose differences between two programs. CoverUp~\cite{pizzorno2024coverup} is another tool that iteratively improves the generated tests by asking the model for new tests that cover uncovered lines. Alshahwan~\etal~\cite{alshahwan2024automated} design four prompts that improve the coverage and corner case detection of generated tests. Steenhoek~\etal~\cite{steenhoek2023reinforcement} use a new reward model to generate tests that follow best practice patterns.

Li~\etal~\cite{li2023nuances} introduce differential prompting, an LLM-based tool that generates fault-inducing tests. This tool takes a buggy program. It first uses the LLM to generate an intention description for the program. Then, it generates a fixed version of the program. Next, it employs the LLM to generate a list of test inputs. Finally, a test input that leads to different results on the buggy and fixed versions is labeled as fault-inducing. AID~\cite{liu2024llm} is another tool that follows the same procedure, while it uses the LLM to generate a test generator, instead of test inputs. These two tools are different from \toolname as they generate fault-inducing tests for a given buggy program. In contrast, \toolname directly generates difference exposing tests for two given programs.

Liu~\etal~\cite{liu2024your} utilize LLMs to amplify existing tests in the HumanEval dataset~\cite{chen2021evaluating}. They generate new tests that expose differences between the dataset programs and incorrect programs generated for the code documentation. Their test generation method starts from test seeds generated by LLMs and then improves them by type-aware mutation. This method is different from \toolname, which uses an execution-driven iterative approach to improve initial tests generated by the LLM.

%\subsection{Generating Tests for Python}
%\label{sec:tests-for-python}

\subsection{Generating Tests for Python}
Generating tests for dynamically typed programming languages is challenging~\cite{selakovic2018test}. Auger~\cite{auger} is one of the first tools to address this task for Python. Auger uses Python tracer to collect the values with which target functions are called and the values that target functions return. The collected data is then used to generate unit tests.

The state-of-the-art tool for Python test generation is Pynguin~\cite{lukasczyk2022pynguin,lukasczyk2020automated,erni2024sbft}, which uses evolutionary algorithms to generate unit tests with high coverage. It uses type annotations and static analysis to infer the type of function parameters. Lukasczyk~\etal~\cite{lukasczyk2023empirical} conduct a study to improve Pynguin's initial type inference strategy and find a more suitable evolutionary algorithm: DynaMOSA~\cite{panichella2017automated}.

Schoofs~\etal~\cite{schoofs2022ampyfier} introduce AmPyfier, which amplifies existing Python test suites. AmPyfier runs existing tests on the PUT and collects values during execution. These collected values are then added to the existing test as new assertions. AmPyfier also amplifies the inputs to PUT iteratively. To make the amplification efficient, at each iteration, AmPyfier only keeps the assertion and input amplifications that increase the mutation or code coverage score.

In contrast with previous Python test generation tools, \toolname focuses on difference-exposing test generation.

%\subsection{Differential Testing}
%\label{sec:differential-testing}

\subsection{Differential Testing}
\modified{Differential testing is a well-established technique for detecting anomalies and defects in programs~\cite{mckeeman1998differential,evans2007differential,taneja2008diffgen,nilizadeh2019diffuzz,rutledge2022automating,ye2023generative,jakobs2022peqtest,groninger2024changeguard}.} It is employed to identify issues in many types of programs, such as robotic vehicle programs~\cite{kim2023patchverif}, game agents~\cite{castellano2022explaining}, spreadsheet applications~\cite{almasi2018search}, certificate parsers~\cite{chen2023sbdt}, and model counters~\cite{usman2020testmc}. Research has shown the usefulness of differential testing in large-scale industrial software projects~\cite{gulzar2019perception,jarman2020program,jarman2020difference}.

Silva~\etal~\cite{da2020detecting} study the feasibility of detecting semantic conflicts at merge time with differential testing. They use EvoSuite~\cite{fraser2011evosuite} and Randoop~\cite{pacheco2007randoop} to generate tests before a branch is merged. They show that the generated tests can reveal semantic conflicts between the base and merging branches.

\modified{Noller et al. \cite{noller2020hydiff} introduce HyDiff, a hybrid approach that generates inputs for two Java bytecode programs to expose differences, including functional differences. HyDiff uses AFL fuzzer \cite{afl} to generate new inputs by mutating existing ones, and also employs symbolic execution to generate inputs that cover various branches. In contrast to HyDiff, \toolname takes the source of two Python programs as input and leverages LLMs for test generation.}

% Ye~\etal~\cite{ye2023generative} propose a novel differential fuzzing approach approach, called ComFuzz, to detect compiler bugs. ComFuzz employs the Distill-BERT~\cite{sanh2019distilbert} language model to complete shortened historical programs by adding new statements. It then compiles these programs with multiple compilers and detects inconsistencies between the results. These incosistencies expose potential bugs in compilers. ComFuzz iteratively improves test programs by using rule-based mutation on previously generated ones. In contrast with ComFuzz, \toolname targets application level programs, not compilers. Moreover, \toolname iteratively improves the initial test by providing the LLM with execution-driven feedback, while ComFuzz uses rule-based mutation to improve the initial test program.

Recently, differential testing has also been used for testing deep learning programs and networks~\cite{zhang2021duo}. Guo~\etal~\cite{guo2018dlfuzz} propose Dlfuzz, a tool that generates rare inputs for DL networks. Dlfuzz uses a mutation algorithm to generate new inputs that activate previously unactivated neurons. Liu~\etal~\cite{liu2023generation} propose a novel technique, called Gandalf, that generates inputs to detect inconsistencies between DL libraries. 
%%Gandalf employs the context-free grammar of DL models and a deep Q-Network to generate valid and diverse inputs.

\toolname is the first LLM-based differential testing tool, which uses an execution-driven and iterative prompting approach.

\section{Conclusion}
\label{sec:conclusion}
In this paper, we introduce \toolname, a novel tool that generates difference exposing tests (DET) employing LLMs with an execution-driven iterative approach. \toolname takes two versions of a program (P \& Q) and an example test that produces the same output on P \& Q. \toolname's outcome is a test input that produces different results on P \& Q. \toolname works with advanced prompting of an LLM: the prompt consists of P \& Q, the example test, the output produced by the example test, and monitoring data collected during the execution of the example test. 
We evaluate \toolname on QuixBugs and a novel benchmark called \mainds, both in Python. Our experiments reveal that \toolname is effective and generates a DET for 81.7\% (1,255/\maindssize) of program pairs in \mainds and 100\% (32/\qbsize) of program pairs in QuixBugs. \toolname significantly outperforms Pynguin~\cite{lukasczyk2022pynguin} and \dpp~\cite{li2023nuances}, the closest, state-of-the-art test generation tools for the considered task.

% For future work, we envision incorporating grammar based fuzzing techniques into \toolname. This would enable \toolname to more accurately specify the type and structure of inputs to be given to the target programs.

\bibliographystyle{elsarticle-num}
\bibliography{references}
\end{document}